\documentclass[reprint,nofootinbib,superscriptaddress,amsmath,amssymb,aps,prd]{revtex4-1}
\usepackage{graphicx}
\usepackage{rotating}
\usepackage{dcolumn}
\usepackage{bm}
\usepackage{color}
\usepackage{mathptmx, textcomp}
\usepackage[latin1]{inputenc}
\usepackage{braket}
\usepackage[colorlinks,linkcolor=magenta,anchorcolor=cyan,citecolor=blue]{hyperref}
\usepackage[multiple]{footmisc}

\def\be{\begin{equation}}
\def\ee{\end{equation}}
\def\ba{\begin{eqnarray}}
\def\ea{\end{eqnarray}}

      \def\p {\pi} \def\a {\alpha}       \def\k {\kappa}        \def\p {\pi} \def \inf {\infty}  
        \def\F {\Phi}  \def\L {\Lambda}    \def\grad{\nabla}\def\.{\cdot}
\def\math {\mathcal}
\begin{document}

\title{Holographic Complexity in Charged Accelerating Black Holes}

\author{Shun Jiang}
\email{shunjiang@mail.bnu.edu.cn}
\author{Jie Jiang}
\email{Corresponding author. jiejiang@mail.bnu.edu.cn}
\affiliation{%
	Department of Physics, Beijing Normal University, Beijing, 100875, China\\}
\affiliation{College of Education for the Future, Beijing Normal University, Zhuhai 519087, China}

\begin{abstract}
Using the ``complexity equals action''(CA) conjecture, for an ordinary charged system, it has been shown that the late-time complexity growth rate is given by a difference between the value of $\Phi_{H}Q+\Omega_H J$ on the inner and outer horizons. In this paper, we investigate the complexity of the boundary quantum system with conical deficits. From the perspective of holography, we consider a charged accelerating black holes which contain conical deficits on the north and south poles in the bulk gravitational theory and evaluate the complexity growth rate using the CA conjecture. As a result, the late-time growth rate of complexity is different from the ordinary charged black holes. It implies that complexity can carry the information of conical deficits on the boundary quantum system.

\end{abstract}
\maketitle

\section{Introduction}\label{section1}
The quantum circuit complexity, defined as the minimum number of elementary gates required to construct a target state from a reference state, has attracted increasing interest in recent years \cite{Susskind:2014rva,Aaronson:2016vto}. From the viewpoint of AdS/CFT, Brown \textit{et al}. suggest that the quantum complexity of state in the boundary theory is corresponding to some bulk gravitational quantities which are called ``holographic complexity''. The two most famous conjectures are ``complexity equals volume'' (CV)\cite{Susskind:2014rva,Stanford:2014jda} and ``complexity equals action'' (CA) \cite{Brown:2015bva,Brown:2015lvg}. The former relates the complexity to the volume of extremal codimension-one surfaces in the bulk. The latter relates the complexity to the value of gravitational action within the wheeler-Dewitt(WDW) patch. These conjectures have aroused extensive researchers' attention to both holographic complexity and circuit complexity in quantum field theory \cite{Roberts:2014isa,Guo:2020dsi,Fan:2018xwf,Fan:2018wnv,Lehner:2016vdi,Jefferson:2017sdb,Carmi:2016wjl,Chapman:2017rqy,Carmi:2017jqz,Caputa:2017yrh,Chapman:2016hwi,Brown:2017jil,Couch:2016exn,Cai:2016xho,Susskind:2014jwa,Roberts:2016hpo,Ben-Ami:2016qex,Khan:2018rzm,Hackl:2018ptj,Chapman:2018dem,Chapman:2018hou,Kim:2017qrq,Brown:2016wib,Reynolds:2016rvl,Agon:2018zso,Chapman:2018lsv,Aaronson:2016vto,Caputa:2018kdj,Abt:2017pmf,Hashimoto:2017fga,Guo:2018kzl,Bhattacharyya:2018wym,Alishahiha:2017hwg,Alishahiha:2018tep,Swingle:2017zcd,Cai:2017sjv,Bhattacharyya:2018bbv,Yang:2017nfn,Belin:2018bpg,An:2018xhv,Fu:2018kcp,Chen:2018mcc,Barbon:2015ria,Czech:2014ppa,Yang:2018nda,Takayanagi:2018pml,Susskind:2015toa,Kim:2017lrw,Cano:2018aqi,Qi:2018bje,Reynolds:2017lwq,Ali:2018fcz,Couch:2017yil,Pan:2016ecg,Alves:2018qfv,Huang:2016fks,Guo:2017rul,Reynolds:2017jfs}

Here we only consider the CA conjecture which asserts the complexity of the CFT state is given by the numerical value of the gravitational action evaluated on the WDW patch:
\begin{eqnarray}
	C_{A}(\left|\Psi \right\rangle )=\frac{I_\text{WDW}}{\pi \hbar}
\end{eqnarray}
By investigating various black holes, some general behaviors have been uncovered. For example, the response of complexity to perturbations follows the ``switchback effect''\cite{Chapman:2018lsv}. Another well understanding behavior is that the late-time complexity grows linearly in time at a rate characterized by the conserved charges and thermodynamic potentials on the inner and outer horizons of the black hole\cite{Brown:2015lvg,Cai:2016xho,Huang:2016fks,Carmi:2017jqz,Cano:2018aqi}. For the charged and rotating black holes with multiple horizons, a series of works \cite{Cai:2016xho,Guo:2017rul,Pan:2016ecg,Jiang:2018pfk} show that the CA complexity grow rate at late times can be expressed as
\begin{eqnarray} \lim_{t\rightarrow\infty}\frac{dC_{A}}{dt}
=\frac{1}{\pi\hbar}[(M-\Phi_HQ-\Omega_HJ)]^{+}_{-} \label{latetime}
\end{eqnarray}
where $Q$ and $J$ are the electric charges and angular momentum of black holes, $\Omega^{(\pm)}_{H}$ and $\Phi^{(\pm)}_{H}$ are the angular velocity associated with the inner and outer horizons respectively and the index $(\pm)$ presents the outer and inner horizon. However, these results are based on the black holes which don't have conical deficits, and therefore, the corresponding boundary systems don't have conical deficits. It is natural to ask whether the complexity can reflect the information of the conical deficits on the boundary quantum system. From the viewpoint of AdS/CFT correspondence, a boundary system with conical deficits is dual to a bulk black hole system with conical deficits. Therefore, in this paper, we consider the charged accelerating black holes which have two conical deficits on the north and south poles. The charged accelerating black hole solutions were obtained in \cite{Plebanski:1976gy,Zhang:2019vpf}.

The remainder of this paper is as follows. In Sec.~\ref{section2}, we review the geometry of charged accelerating black holes and some thermodynamical quantities. In Sec.~\ref{section3}, we evaluate the late-time result of the complexity growth rate using the CA conjecture. In Sec.~\ref{section4}, we give a brief conclusion.

\section{Geometry of charged accelerating black hole}\label{section2}
In this paper, we consider the four-dimensional charged accelerating black holes with the bulk action
\begin{eqnarray}
I_\text{bulk}=\frac{1}{16\pi}\int_{M}\bm\epsilon(R-2\Lambda-\math{F})\,,
\end{eqnarray}
with $\math{F}=F_{ab}F^{ab}$,  in which $R$ is the Ricci scalar, $\L$ is cosmological constant, and $\bm{F}=d\bm{B}$ with electromagnetic filed $\bm{B}$ is the electromagnetic strength. We consider the solutions which have maximally symmetries. It means the curvature scalar $R$ is a constant, i.e., $R=R_0=4\L$($R_0<0$ for AdS black hole). The corresponding equations of motion reads
\ba\begin{aligned}
&R_{ab}-\frac{1}{4}R_0g_{ab}=2T_{ab}\,,\quad\quad\grad_a F^{ab}=0
\end{aligned}\label{motion}\ea
with
\begin{eqnarray}
T_{ab}=F_{a}{}^{c}F_{bc}-\frac{1}{4}\math{F}g_{ab}\,.
\end{eqnarray}
According to Eqs.~(\ref{motion}), the line element of the charged accelerating black hole solution reads
\ba\begin{aligned} ds^2=&\frac{1}{\Omega^2}\left[-\frac{N(r)}{\alpha^2}dt^2+\frac{1}{N(r)}dr^2\right.\\
&\left.+r^2\left(\frac{d\theta^2}{H(\theta)}+H(\theta)\sin^2\theta\frac{d\phi^2}{k^2}\right)\right]\,, \label{metric}
\end{aligned}\ea
where
\ba\begin{aligned}
&\Omega=1+Ar\cos\theta\,,\\ &N(r)=(1-A^2r^2)\left(1-\frac{2m}{r}+\frac{q^2}{r^2}\right)-\frac{R_0r^2}{12}\,,\\
&H(\theta)=1+2mA\cos\theta+q^2A^2\cos^2\theta\,,\\
&\alpha=\sqrt{\Xi\left(1+\frac{12A^2\Xi}{R_0}\right)}\,,\quad \Xi=1+q^2A^2\,.\\
\end{aligned}\ea
In this black hole solution, the conformal factor $\Omega$ determines the conformal boundary 
\ba
r_b=-\frac{1}{A\cos\theta}\,.
\ea 
Note that $A$, $m$ and $q$ are the acceleration, mass and electric parameters of the black hole. $k$ stands for the conical deficits of the black hole on the north and south poles. The parameter $\alpha$ is used to rescale the time coordinate so that one can get a normalized Killing vector at conformal infinity. Solving Eqs.~(\ref{motion}) for gauge field, we get
\begin{eqnarray}
	B_a=-\frac{q}{\a r}(dt)_a\,.
\end{eqnarray}
Thermodynamics of the charged accelerating black hole has been studied in Ref. \cite{Zhang:2019vpf}. The mass of charged accelerating black hole is given by
\begin{eqnarray}
	M=\frac{m(1-A^2l^2\Xi)}{k\alpha}
\end{eqnarray}
We are considering charged accelerating black holes in Einstein gravity, the entropy $S$ of charged accelerating black hole can be written as
\begin{eqnarray}
	S=\frac{A_H}{4}
\end{eqnarray}
where $A_H$ is the area of the event horizon in the charged accelerating black hole. The temperature of the event horizon can be obtained as
\begin{eqnarray}
	T=\frac{f'(r_{+})}{4\pi\alpha}
\end{eqnarray}
The electric charge of charged accelerating black can be written as
\begin{eqnarray}
	Q=\frac{1}{4\pi}\int\star\boldsymbol{F}
\end{eqnarray}
The conjugate electric potential of the event horizon is
\begin{eqnarray}
	\Phi_H^{(+)}=\frac{q}{\alpha r_{+}}
\end{eqnarray}

\section{Complexity Growth Rate In CA Conjecture}\label{section3}
In this section, we evaluate the growth rate of the holographic complexity for the charged accelerating black holes based on the CA conjecture. It means we need to evaluate the on-shell action within the WDW patch. To make the variational principle well-posed, the total action should include boundary terms, joint terms, and counterterms. Therefore, the total action can be written as
\ba\begin{aligned} I_\text{total}=&I_\text{bulk}+\frac{1}{8\pi}\int_{B}d^3x\sqrt{|h|}K+\frac{1}{8\pi}\int_{S}d^2\theta\sqrt{\sigma}\eta\\
&+\frac{1}{8\pi}\int_{N}d\lambda d^2\theta\sqrt{\gamma}\kappa+\frac{1}{8\pi}\int_{N}d\lambda d^2\theta\sqrt{\gamma}\Theta \ln(l_\text{ct}\Theta)\,,
\end{aligned}\ea
where $B$ and $N$ are the non-null and null segments of the boundary of the WDW patch, and $S$ is a two-dimensional joint of the nonsmooth boundary. Here, $h_{ab}$ and $K$ are the induced metric and trace of the extrinsic curvature, $\gamma_{ab}$ is the induced metric on the cross section of the null segment $N$, $\lambda$ is the parameter of the null generator $k^a$ on the null segment, the parameter $\kappa$ is given by $k^a\nabla_{a} k^b=\kappa k^b$ and it measures the failure of $\lambda$ to be an affine parameter, $\Theta=\nabla_ak^a$ is the expansion scalar of the null generator, and $l_\text{ct}$ is some arbitrary constant parameter.
\begin{figure}
	\centering
	\includegraphics[scale=1]{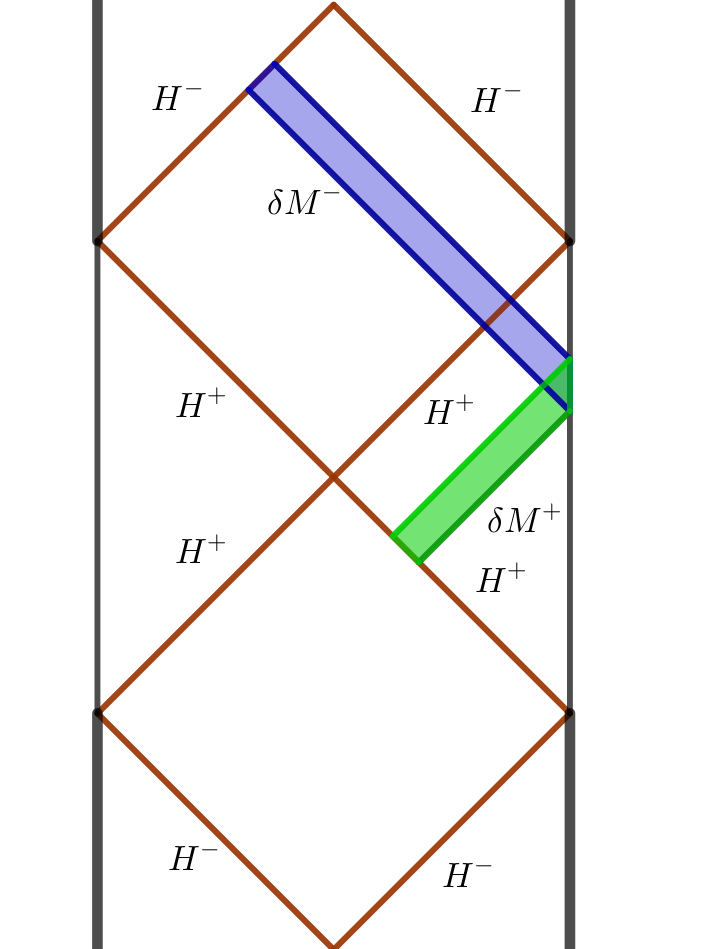}
	\caption{A spacetime diagram of charged accelerating black holes with two killing horizons. $H^{+}$ and $H^{-}$ are the corresponding outer and inner horizons. We show the change of the Wheeler-DeWitt patch, in which we fix the left boundary at $t_{L}\to \infty$ and vary the right boundary}
	\label{fig1}
\end{figure}

In Fig.\ref{fig1}, we show the Penrose diagram of charged accelerating black holes which has two killing horizons. The spacetime is invariant under shift transformation $t_R\rightarrow t_R-\delta t$, $t_L\rightarrow t_L+\delta t$. Therefore, we can fix the left boundary time $t_L=\infty$ and vary the right boundary time $t_R$, i.e., we only need to evaluate the action within the regions  $\delta M^{(\pm)}$ to show the late-time complexity growth rate. 

\subsection{Bulk Contributions}
In this subsection, we calculate the contributions from bulk action. This corresponds to bulk region $\delta M^{(\pm)}$. Since we can use the same technology for $\delta M^{(\pm)}$, we will neglect the index $(\pm)$. From Eqs.~(\ref{motion}), we have
\begin{eqnarray} R_{ab}-\frac{1}{4}R_0g_{ab}=2F_{a}{}^{c}F_{bc}-\frac{1}{2}\math{F}g_{ab}
\end{eqnarray}
The bulk action within region $\delta M^{(\pm)}$ can be written as
\ba\begin{aligned}
I_{\delta M}&=\frac{1}{16\pi}\int_{\delta M} \star\left(\frac{1}{2}R_0-\math{F}\right)\\
&=\frac{1}{16\pi} \delta t \int_N\left(\frac{1}{2}R_0-\math{F}\right) \star\bm k\,,
\end{aligned}\label{bulk}\ea
in which $k^a_{(\pm)}$ is the Killing vector filed of the inner and outer horizons.  Considering a killing vector field $k^a$ and using the killing equation $\nabla_a k_b+\nabla_b k_a=0$, we find
\ba\begin{aligned}
\nabla_a\nabla^a k^b&=-R^b{}_a k^a\\
&=-\frac{1}{4}R_0 k^b-2T^b{}_a k^a\\
&=-\frac{1}{4}R_0 k^b-2F^{bc}F_{ac}k^a+\frac{1}{2}\math{F} k^b\,.
\end{aligned}\ea
The second term can be written as
\ba\begin{aligned}
k^aF_{ac}F^{bc}&=k^a(dA)_{ac}F^{bc}\\
&=F^{bc}\mathcal{L}_k A_c+F^{bc}\nabla_c\Phi\\
&=-\nabla_a(\Phi F^{ab})+\Phi \nabla_a F^{ab}
\end{aligned}\ea
with
\ba\begin{aligned}
	\Phi=-A_a k^a\,.
\end{aligned}\ea
Thus, we have
\ba\begin{aligned}
	2\nabla_a\nabla^a k^b=-\frac{1}{2}R_0k^b+4\nabla_c(\Phi F^{cb})+\math{F}k^b\,. \label{kequation}
\end{aligned}\ea
Using Eqs.~(\ref{kequation}), Eq.~(\ref{bulk}) can be written as
\ba\begin{aligned} \left(\frac{1}{2}R_0-\math{F}\right)\star\boldsymbol{k}=d\star\boldsymbol{k}-4d\star(\Phi\boldsymbol{F})\,.
\end{aligned}\ea
Using the above result, we obtain
\ba\begin{aligned}
&\frac{16\p I_{\delta M}}{\delta t}=\int_{N}d\star(d\boldsymbol{k}-4\Phi \boldsymbol{F})\\
&=\int_{r_{\infty}}\star d\boldsymbol{k}-\int_{r_h}\star d\boldsymbol{k}-4\int_{r_{\infty}}\Phi\star\bm F+4\int_{r_h}\Phi\star\bm F\\
&-\int_{\theta=\epsilon}[\star d\boldsymbol{k}-4\star(\Phi \boldsymbol{F})]+\int_{\theta=\p-\epsilon}[\star d\boldsymbol{k}-4\star(\Phi \boldsymbol{F})]\,. \label{bulk change}
\end{aligned}\ea
It is worth noting that there are two extra terms in the last line of Eqs.~(\ref{bulk change}). This is because the metric has two conical deficits at the north pole ($\theta\rightarrow0$) and the south pole ($\theta\rightarrow\pi$). When $\theta\rightarrow0$ or $\theta\rightarrow\pi$, the trace of extrinsic curvature $K\rightarrow\infty$. Therefore, it is nature to consider two boundary surface located at $\theta=\epsilon$ and $\theta=\p-\epsilon$, where $\epsilon\rightarrow0$. Then, it is not difficulty to show
\ba\begin{aligned}
	\frac{1}{16\pi}\int_{\theta=\epsilon}\star d\boldsymbol{k}&\sim \sin^2\epsilon\\
\frac{1}{4\pi}\int_{\theta=\epsilon}\star(\Phi \boldsymbol{F})&\sim \sin^2\epsilon\,.
\end{aligned}\ea
Therefore, conical deficits won't affect the bulk action's growth rate.

Using the definition of the black hole entropy, we have
\ba\begin{aligned}
-\frac{1}{16\p}\int_{r_h} \star d\bm{k}=T S\,.
\end{aligned}\ea
Considering  $\F(r_h)=\F_H$ is a constant on the horizon, using the definition of electric charge, we have
\ba\begin{aligned}
Q=\frac{1}{4\pi}\int \star\bm{F}\,.
\end{aligned}\ea
Combing the above result, the late-time growth rate of bulk action becomes
\ba\begin{aligned}
	\lim_{t\to \inf}\frac{d I_\text{bulk}}{d t}=\left[\Phi_H Q+T S\right]_{+}^{-}\,.
\end{aligned}\label{latebulk}\ea
\subsection{Joint Contributions}\label{jonitC}
In this subsection, we will calculate the contribution from the joint term. Without loss of generality, we choose the affine parameter for the null generator $l^a$ of the right null surface. Meanwhile, we require $l^a$ satisfies $\mathcal{L}_k l^a=0$. We first focus on the joint point $r=r_{-}$ which is the intersection point of the inner horizon. The inner horizon is generated by the killing vector $k^a$ and it satisfies
\ba
k^b\nabla_b k^a=\kappa k^a\,.
\ea
Therefore, the affinely null generator on the horizon can be constructed as $\bar{k}^a=e^{-\k t}k^a=e^{-\k t}(\frac{\partial}{\partial t})^a$. The transformation parameter can be written as
\ba\begin{aligned}
	\eta(t)=\ln\left(-\frac{1}{2}\bar{k}\cdot l\right)=-\kappa t+\ln\left(-\frac{1}{2}k\cdot l\right)\,.
\end{aligned}\ea
Using the above relation, the joint contribution at $r=r_{-}$ can be written as
\ba\begin{aligned} \frac{dI_\text{joint}}{dt}=\frac{1}{8\pi}\frac{d}{dt}\int_{S}d^2\theta\sqrt{\sigma}\eta=-\frac{\kappa}{2\pi}S=-T^{(-)}S^{(-)}\,.
\end{aligned}\ea
A similar calculation shows the contribution from past meeting point $r=r_{+}$ has the same form. The late-time growth rate of joint contribution is given by
\ba\begin{aligned} \lim_{t\rightarrow\infty}\frac{dI_\text{joint}}{dt}=T^{(+)}S^{(+)}-T^{(-)}S^{(-)}\,.
\end{aligned}\ea
From Eq.~(\ref{latebulk}), we see this term will cancel $TS$ term in bulk contribution.
\subsection{Counterterm Contributions}
In this subsection, we will evaluate the counterterm contributions. The counterterm contributions can be written as
\ba\begin{aligned}
	I_\text{ct}=\frac{1}{8\pi}\int_{N}d\lambda d^2\theta\sqrt{\gamma}\Theta \ln(l_\text{ct}\Theta)
\end{aligned}\ea
We first consider the upper two null segements. The right null segement is generated by $l^a$. The left null segement is on inner horizon and we choose the affinely null generator $\bar{k}^a=e^{-kt}k^a$. In previous subsection, we require the the null generator $l^a$ satisfies $\mathcal{L}_k l^a=0$. It is easy to see $\mathcal{L}_k\Theta=0$ for both generators. Therefore, the rate of counterterm contributions vanish.

\subsection{Surfaces Contributions}
From Eq.~(\ref{metric}), we see the metric has two conical deficits at the north pole ($\theta\rightarrow0$) and south pole ($\theta\rightarrow\pi$). The trace of extrinsic curvature $K\rightarrow\infty$ when
$\theta\rightarrow0$ and $\theta\rightarrow\pi$. Therefore, we should add two extra boundary surfaces at $\theta=\epsilon$ and $\theta=\pi-\epsilon$. The surfaces contributions from $\theta\rightarrow0$ and $\theta\rightarrow\pi$ is given by
\ba\begin{aligned} I_{\Sigma_\theta}=\frac{\text{sign}(\sigma)}{8\pi}\int_{\Sigma_\theta}\boldsymbol{\varpi}\,, \label{surface}
\end{aligned}\ea
where $\text{sign}(\sigma)=-1$ for $\theta=\epsilon$, $\text{sign}(\sigma)=-1$ for $\theta=\pi-\epsilon$ and three form $\boldsymbol{\varpi}$ can be written as
\ba\begin{aligned}
	\boldsymbol{\varpi}=\sqrt{|h|}Kdt\wedge dr \wedge d\phi\,.
\end{aligned}\ea
Where $h$ is the determinant of induced metric.

Using the language of differential forms, the above Eq.~(\ref{surface}) can be expressed as
\ba\begin{aligned}
I_{\Sigma_\theta}=\delta t\frac{\text{sign}(\sigma)}{8\pi}\int_{\Sigma_\theta}\xi\cdot\boldsymbol{\varpi}\,.
\end{aligned}\ea
From Eq.~(\ref{metric}), we find
\begin{eqnarray}
	I_{\Sigma_\theta}=\frac{\text{sign}(\sigma)\delta t}{8\pi} \int \gamma^{(\pm)}\frac{1}{(1\pm Ar)^2}drd \varphi
\end{eqnarray}
where $\gamma^{(\pm)}$ is given by
\begin{eqnarray} \gamma^{(\pm)}=-\sqrt{\frac{(1\pm2Am+A^2q^2)(1+2Am+A^2q^2)R_0}{k^2(1+A^2q^2)(12A^2+12A^4q^2+R_0)}}\,.
\end{eqnarray}
Then, at late times, the surfaces contributions can be written as
\ba\begin{aligned}
	I_{\Sigma_\theta}=&\frac{\gamma^{(+)}\delta t}{4\pi A}\left(\frac{1}{1+Ar_{-}}-\frac{1}{1+Ar_{+}}\right)\\
&+\frac{\gamma^{(-)}\delta t}{4\pi A}\left(\frac{1}{1-Ar_{+}}-\frac{1}{1-Ar_{-}}\right)
\end{aligned}\ea
\subsection{Complexity growth rate}
Combining all the previous results, the late-time complexity growth rate is given by
\ba\begin{aligned}	\frac{1}{\p\hbar}\frac{dC_{A}}{dt}=&\left(\Phi^{(-)}_H-\Phi^{(+)}_H\right)Q+\frac{\gamma^{(+)}}{4\pi A}\left(\frac{1}{1+Ar_{-}}-\frac{1}{1+Ar_{+}}\right)\\
&+\frac{\gamma^{(-)}}{4\pi A}\left(\frac{1}{1-Ar_{+}}-\frac{1}{1-Ar_{-}}\right)\,.
\end{aligned}\ea
This result is different from the ordinary charged systems. For the ordinary charged systems, it has been shown that the late-time growth rate is $\Phi_H^{(-)} Q-\Phi_H^{(+)}Q$. Because of the conical deficits, there are some extra terms that are evaluated on the inner and outer horizons.

\section{Conclusion}\label{section4}
In this paper, we considered the charged accelerating black holes which have two conical deficits on the north and south poles in Einstein-Maxwell theory. From the perspective of AdS/CFT, the dual bound system of this bulk gravity also has two conical deficits on the north and south poles. To investigate the influence of the conical deficits on the complexity of the boundary system, we evaluated the growth rate of the  CA complexity in charged accelerating black holes. We show that the time-time complexity growth rate is different from the result Eq.~(\ref{latetime}) of an ordinary charged system. There is an additional term that is evaluated on the inner and outer horizons. These imply that the complexity can reflect some information of conical deficits in the boundary CFT system.

\section{Acknowledgments}
This research was supported by NSFC Grants No. 11775022 and 11873044.

\end{document}